\documentclass{iau}
\usepackage{graphicx}

\title[Revealing obscured AGN with WISE] 
{Effectiveness of WISE colour-based selection techniques to uncover obscured AGN}

\author[S. Mateos]   
{S. Mateos$^1$
}

\affiliation{$^1$Instituto de F\'isica de Cantabria (CSIC-Universidad de Cantabria), 39005 Santander, Spain \\ email: {\tt mateos@ifca.unican.es} \\[\affilskip]
}

\pubyear{2013}
\volume{304}  
\pagerange{119--126}
\jname{Multiwavelegth AGN Surveys and Studies}
\editors{A. Mickaelian, F. Aharonian, D. Sanders, eds.}
\begin{document}

\maketitle

\begin{abstract}
We present a highly reliable and efficient mid-infrared colour-based
selection technique for luminous active galactic nuclei (AGN) using
the Wide-field Infrared Survey Explorer (WISE) survey. Our technique
is designed to identify objects with red mid-infrared power-law
spectral energy distributions. We studied the dependency of our
mid-infrared selection on the AGN intrinsic luminosity and the
effectiveness of our technique to uncover obscured AGN missed in X-ray
surveys. To do so we used two samples of luminous AGN independently
selected in hard X-ray and optical surveys. We used the largest
catalogue of 887 [${\rm O_{III}}$] $\lambda$5007-selected type 2
quasars (QSO2s) at z$\lesssim$0.83 in the literature from the Sloan
Digital Sky Survey (SDSS), and the 258 hard ($>$4.5 keV)
X-ray-selected AGN from the Bright Ultrahard XMM-Newton Survey
(BUXS). The effectiveness of our mid-infrared selection technique
increases with the AGN luminosity. At high luminosities and at least
up to $z$$\sim$1 our technique is very effective at identifying both
Compton-thin and Compton-thick AGN.

\end{abstract}

\firstsection 
\section{Introduction}
A complete census of the obscured AGN is crucial to fully understand
the cosmological growth of supermassive black holes (SMBH) and to
reveal the nature of the SMBH-galaxy co-evolution. Obscured accretion
is a key phase both in AGN growth and in the co-evolution of AGN and
their host galaxies as most SMBH mass growth occurs in heavily
obscured environments (\cite[Fabian \& Iwasawa
  1999]{1999MNRAS.303L..34F}). However, even the deepest X-ray surveys
conducted to date with XMM-Newton and Chandra at energies $>$2 keV are
incomplete for AGN with line-of-sight neutral hydrogen column
densities ${\rm N_H > 10^{23}\,cm^{-2}}$ and they miss the
Compton-thick AGN (${\rm N_H > 1.5\times10^{24}\,cm^{-2}}$;
e.g. \cite[Burlon et al. 2011]{2011ApJ...728...58B}).

Surveys at mid-infrared wavelengths ($>$5$\mu$m) are much less
affected by extinction since the obscuring circumnuclear dust
reradiates the absorbed nuclear optical-to-X-ray radiation in the
infrared. As shown by selection techniques mainly developed with data
from Spitzer-IRAC, surveys conducted at mid-infrared wavelengths can
potentially reveal the elusive obscured accretion missed by hard X-ray
surveys (e.g. \cite[Lacy et al. 2004]{lacy04}; \cite[Stern
  et al. 2005]{stern05}; \cite[Alonso-Herrero et al. 2006]{alonso06};
\cite[Donley et al. 2012]{2012ApJ...748..142D}). AGN population
studies with the Wide-field Infrared Survey Explorer (WISE;
\cite[Wright et al. 2010]{2010AJ....140.1868W}) are starting to fill
the gap between local/deep mid-infrared surveys with IRAS/Spitzer,
completing our census of obscured SMBH growth in regions of the
luminosity-redshift parameter space poorly sampled. Several works have
already demonstrated that using WISE colours alone it is possible to
separate stars and star-forming galaxies from luminous AGN
(e.g. \cite[Mateos et al. 2012]{mateos12}; \cite[Stern
  et al. 2012]{stern12}; \cite[Assef et al. 2013]{assef13};
\cite[Mateos et al. 2013]{2013MNRAS.434..941M}; \cite[Yan
  et al. 2013]{2013AJ....145...55Y}).

In \cite[Mateos et al. (2012)]{mateos12} (hereafter M12) we presented a
colour-based selection technique of luminous AGN using the 3.4, 4.6,
and 12 $\mu$m bands of WISE (hereafter mid-infrared wedge). We
demonstrated this technique is one of the most reliable and efficient
to detect X-ray luminous AGN in the literature. Furthermore, in
\cite[Mateos et al. (2013)]{2013MNRAS.434..941M} (hereafter M13) we
showed it is very effective at revealing both Compton-thin and
Compton-thick luminous AGN, at least up to z$\lesssim$1. Here we
briefly summarise the results of these studies.

\section{The data}
The samples used are described in detail in M12, M13 and \cite[Reyes
  et al. (2008)]{2008AJ....136.2373R} (hereafter R08). Briefly, the
Bright Ultra-hard XMM-Newton Survey (BUXS) is a complete flux-limited
sample of 258 bright (${\rm f_{4.5-10\,keV} > 6 \times
  10^{-14}\,erg\,s^{-1}\,cm^{-2}}$) “ultra-hard” (4.5-10 keV) X-ray
selected sources detected over a sky area of 44.43 deg${\rm ^2}$.
Currently 253 objects have been identified through optical
spectroscopy, 143 as type 1 AGN (Sy1-1.5) and 110 as type 2 AGN
(Sy1.8-2). 

We also have used the largest catalogue of 887 [${\rm O_{III}}$]
$\lambda$5007-selected type 2 quasars (QSO2s) at $z$$\lesssim$0.83 in
the literature from the Sloan Digital Sky Survey (SDSS) from
R08. These QSO2s were selected independently of their X-ray properties
hence, the sample should be unaffected by nuclear obscuration.

\begin{figure}[t]
 \vspace*{-0.3 cm}
\begin{center}
 \includegraphics[angle=0,width=3.0in]{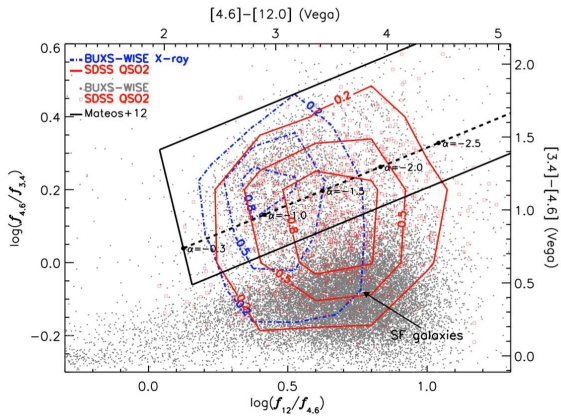} 
 \vspace*{-0.3 cm}
 \caption{Mid-infrared colours for the WISE sources detected in the
   BUXS area down to the full depth of the X-ray observations
   (i.e. not applying the X-ray flux limit that defines the BUXS AGN;
   filled circles).  Open squares are the SDSS QSO2s in R08. The M12
   mid-infrared wedge and power-law locus (and the values for different
   spectral index) are the thick solid and dashed lines,
   respectively. The solid and dot-dashed contours indicate the
   densities (normalized to the peak value) of the SDSS QSO2s and the
   WISE sources in BUXS detected in X-rays, respectively.}
   \label{fig1}
\end{center}
\end{figure}

\section{Results}
\subsection{Mid-infrared selection of AGN with WISE}
Fig.\,\ref{fig1} illustrates our mid-infrared wedge and power-law
locus and the WISE colour distributions of the SDSS QSO2s. For
comparison we show the colours of all the WISE sources detected in the
BUXS fields and of those detected at 2-10 keV energies down to the
full depth of the XMM-Newton observations. The great majority of WISE
objects detected in X-rays fall in the mid-infrared wedge and cluster
near the power-law locus. It seems however, that a substantial
fraction of the SDSS QSO2s have the infrared colours of low $z$
star-forming galaxies (horizontal sequence in the lower-right part of
the diagram).

\subsection{Dependence on AGN luminosity}
\begin{figure}[t]
 \vspace*{0.2 cm}
\begin{center}
 \includegraphics[angle=90,width=2.7in]{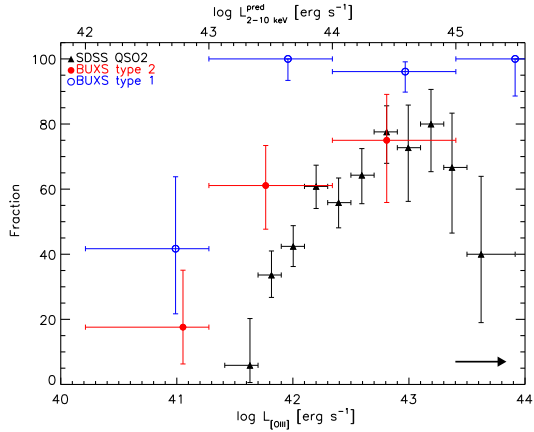} 
 \vspace*{0.3 cm}
 \caption{Fraction of sources in the mid-infrared wedge as a function
   of the AGN luminosity. Triangles are the SDSS QSO2s and open and
   filled circles are the type 1 and type 2 AGN in the BUXS survey,
   respectively. At ${\rm L_{2-10\,keV}>10^{44}\,erg\,s^{-1}}$ $>$96\%
   and $>$75\% of the BUXS type 1 and type 2 AGN and $>$66\% of the
   SDSS QSO2s fall in the mid-infrared wedge, respectively. The
   horizontal arrow at the bottom right shows the amplitude of the
   median extinction correction to the [${\rm O_{III}}$] line
   luminosities.}
   \label{fig2}
\end{center}
\end{figure}

It is well known that the effectiveness of any mid-infrared selection
technique is a strong function of the AGN luminosity (e.g. M12;
\cite[Donley et al. 2012]{2012ApJ...748..142D}; \cite[Messias et
  al. 2013]{2013arXiv1312.3336M}; M13). Fig.\,\ref{fig2} shows the
dependence on luminosity of the fraction of SDSS QSO2s and BUXS AGN in
the mid-infrared wedge. As most SDSS QSO2s have not been observed in
X-rays to compare with the results for the AGN in BUXS we derived
their intrinsic 2-10 keV luminosities using the empirical relation
between hard X-ray emission and [${\rm O_{III}}$] $\lambda$5007
luminosity from \cite[Jin, Ward, \& Done (2012)]{2012MNRAS.422.3268J}
(top axis in Fig.\,\ref{fig2}). We see that the effectiveness of our
selection technique increases with the AGN luminosity. The fraction of
SDSS QSO2s in the mid-infrared wedge is substantially lower than for
the BUXS type 1 AGN but it is consistent, within the uncertainties,
with that for the BUXS type 2 AGN, especially at the highest
luminosities. The apparent different fractions of type 1 and type 2
AGN in the mid-infrared wedge could be explained if all type 2 objects
suffer larger extinction at rest-frame near-infrared wavelengths so
that, for a given luminosity, their observed WISE fluxes, especially
at the shortest wavelengths, are more contaminated by their host
galaxies than those of type 1 AGN. Still, at luminosities ${\rm
  L_{2-10\,keV} > 10^{44}\,erg\,s^{-1}}$ the mid-infrared wedge is
highly effective at selecting obscured AGN ($75.0_{-19.1}^{+14.1}\%$
for the BUXS type 2 AGN and $66.1_{-4.7}^{+4.5}\%$ for the SDSS
QSO2s).

\subsection{Effectiveness of mid-infrared selection to uncover Compton-thick AGN}
\begin{figure}[t]
 \vspace*{0.3 cm}
\begin{center}
 \includegraphics[angle=90,width=2.7in]{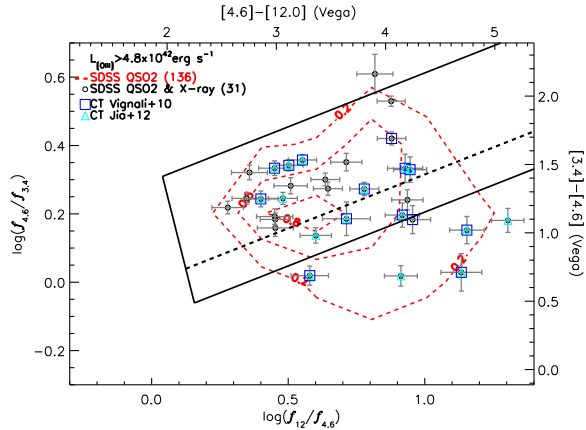} 
 \vspace*{0.2 cm}
 \caption{Mid-infrared colours for the SDSS QSO2s with ${\rm L_{[O_{III}]}
     \gtrsim 4.8 \times 10^{42}\,erg\,s^{-1}}$ and X-ray follow-up
   with either Chandra or XMM-Newton (open circles). Open squares and
   triangles are the Compton-thick candidates from the studies of
   \cite[Vignali et al. (2010)]{vignali10} and \cite[Jia et
     al. (2013)]{2013ApJ...777...27J}, respectively. The mid-infrared
   wedge and power-law locus (and the values for different spectral
   index) are the thick solid and dashed black lines,
   respectively. The dashed contours indicate the density (normalized
   to the peak value) of all SDSS QSO2s with ${\rm L_{[O_{III}]}
     \gtrsim 4.8 \times 10^{42}\,erg\,s^{-1}}$.}
   \label{fig3}
\end{center}
\end{figure}

To investigate the effectiveness of our mid-infrared wedge at
identifying absorbed luminous AGN missed by deep X-ray surveys we have
evaluated whether the SDSS QSO2s identified as Compton-thick
candidates in the literature from the studies of \cite[Vignali et
  al. (2010)]{vignali10} and \cite[Jia et
  al. (2013)]{2013ApJ...777...27J} would be selected by our
mid-infrared wedge. To date the X-ray follow-up of the SDSS QSO2s in
R08 has focused mainly on the most luminous objects hence, in what
follows, we used only the SDSS QSO2s with ${\rm
  L_{[OIII]}>4.8\times10^{42}\,erg\,s^{-1}}$. The fraction of those
objects in the mid-infrared wedge is ${\rm 72.8^{+5.9}_{-6.5}\%}$ (99
out of 136 objects). Out of the 31 SDSS QSO2s in this sample with
X-ray follow-up, 18 objects are robust Compton-thick candidates. Of
these 12 are in the mid-infrared wedge (${\rm
  66.7^{+15.5}_{-18.5}}$\%). We show in Fig.\,\ref{fig3} the WISE
colours of the SDSS QSO2s with X-ray follow-up and indicate those
objects identified as Compton-thick candidates. We see that the
Compton-thick AGN have a distribution of colours that is consistent
with that for the SDSS QSO2s with same luminosities. All these results
fully support that at high AGN luminosities and at least up to
z$\lesssim$1 our mid-infrared selection technique is very effective at
identifying both Compton-thin and Compton-thick AGN.

Based on observations collected at the European Organisation for
Astronomical Research in the Southern Hemisphere, Chile. Based on
observations made with the William Herschel Telescope, the Telescopio
Nazionale Galileo and the Gran Telescopio de Canarias installed in the
Spanish Observatorio del Roque de los Muchachos of the Instituto de
Astrof\'isica de Canarias, in La Palma, Spain. SM acknowledges
financial support from the Spanish Plan Nacional through grants
AYA2010-21490-C02-01 and AYA2012-31447.


\begin{thebibliography}{}

\bibitem[Alonso-Herrero et al. (2006)]{alonso06} {Alonso-Herrero A., et al.} 2006, \textit{ApJ}, 640, 167 

\bibitem[Assef et al. (2013)]{assef13} {Assef R.~J., et al.} 2013, \textit{ApJ}, 772, 26A

\bibitem[Burlon et al. 2011)]{2011ApJ...728...58B} {Burlon D., Ajello M., Greiner J., Comastri A., Merloni A., Gehrels N.} 2011, \textit{ApJ}, 728, 58 

\bibitem[Donley et al. (2012)]{2012ApJ...748..142D} {Donley J.~L., et al. } 2012, \textit{ApJ}, 748, 142 

\bibitem[Fabian \& Iwasawa (1999)]{1999MNRAS.303L..34F} {Fabian A.~C., Iwasawa K.} 1999, \textit{MNRAS}, 303, L34 

\bibitem[Jia et al. (2013)]{2013ApJ...777...27J} {Jia J., Ptak A., Heckman T., Zakamska N.~L.} 2013, \textit{ApJ}, 777, 27 

\bibitem[Jin, Ward, \& Done (2012)]{2012MNRAS.422.3268J} {Jin C., Ward M., Done C.} 2012, \textit{MNRAS}, 422, 3268 

\bibitem[Lacy et al. (2004)]{lacy04} {Lacy M., et al. } 2004, \textit{ApJS}, 154, 166 

\bibitem[Mateos et al. (2012)]{mateos12} {Mateos S., et al. } 2012, \textit{MNRAS}, 426, 3271

\bibitem[Mateos et al. (2013)]{2013MNRAS.434..941M} {Mateos S., Alonso-Herrero A., Carrera F.~J., Blain A., Severgnini P., Caccianiga A., Ruiz A.} 2013, \textit{MNRAS}, 434, 941 

\bibitem[Messias et al. (2013)]{2013arXiv1312.3336M} {Messias H., Afonso J.~M., Salvato M., Mobasher B., Hopkins A.~M.} 2013, \textit{A\&A}, in press (arXiv:1312.3336) 

\bibitem[Reyes et al. (2008)]{2008AJ....136.2373R} {Reyes R., et al. } 2008, \textit{AJ}, 136, 2373 

\bibitem[Stern et al. (2005)]{stern05} {Stern D., et al. } 2005, \textit{ApJ}, 631, 163 

\bibitem[Stern et al. (2012)]{stern12} {Stern D., Assef R.J., Benford D.J., et al. } 2012, \textit{ApJ}, 753, 30

\bibitem[Vignali et al. (2010)]{vignali10} {Vignali C., Alexander D.~M., Gilli R., Pozzi F.} 2010, \textit{MNRAS}, 404, 48

\bibitem[Wright et al. (2010)]{2010AJ....140.1868W} {Wright E.~L., et al. } 2010, \textit{AJ}, 140, 1868 

\bibitem[Yan et al. (2013)]{2013AJ....145...55Y} {Yan L., et al. } 2013, \textit{AJ}, 145, 55 

\end{thebibliography}
\end{document}